\pdfoutput=1
\documentclass[12pt,a4paper]{article}

\usepackage{ifthen} 
\newboolean{pdflatex}
\setboolean{pdflatex}{true} 

\newboolean{articletitles}
\setboolean{articletitles}{true} 

\newboolean{uprightparticles}
\setboolean{uprightparticles}{false} 

\usepackage{bigints}
\def\paperauthors{LHCb collaboration} 
\def\paperasciititle{Test of lepton flavor universality with $B^0 \to K^{*0} \ell^+ \ell^-$ decays at large dilepton invariant mass} 
\def\papertitle{Test of lepton flavor universality with $B^0 \to K^{*0} \ell^+ \ell^-$ decays at large dilepton invariant mass} 
\def\paperkeywords{{High Energy Physics}, {LHCb}} 
\def\papercopyright{\the\year\ CERN for the benefit of the LHCb collaboration} 
\def\paperlicence{CC BY 4.0 licence}
\def\paperlicenceurl{https://creativecommons.org/licenses/by/4.0/}

\newif\ifEnableSectionTOCLinks
\EnableSectionTOCLinksfalse 

\usepackage[top=1in, bottom=1.25in, left=1in, right=1in]{geometry}

\usepackage{microtype}
\usepackage{lineno}  
\usepackage{xspace} 
\usepackage{caption} 

\usepackage[separate-uncertainty = true]{siunitx}

\DeclareSIUnit\clight{\text{\ensuremath{c}}}

\usepackage{graphicx}  
\usepackage{color}
\usepackage{colortbl}
\graphicspath{{./figs/}} 

\usepackage{amsmath} 
\usepackage{amssymb}
\usepackage{amsfonts}
\usepackage{upgreek} 

\newcommand*\patchAmsMathEnvironmentForLineno[1]{%
\expandafter\let\csname old#1\expandafter\endcsname\csname #1\endcsname
\expandafter\let\csname oldend#1\expandafter\endcsname\csname
end#1\endcsname
 \renewenvironment{#1}%
   {\linenomath\csname old#1\endcsname}%
   {\csname oldend#1\endcsname\endlinenomath}%
}
\newcommand*\patchBothAmsMathEnvironmentsForLineno[1]{%
  \patchAmsMathEnvironmentForLineno{#1}%
  \patchAmsMathEnvironmentForLineno{#1*}%
}
\AtBeginDocument{%
\patchBothAmsMathEnvironmentsForLineno{equation}%
\patchBothAmsMathEnvironmentsForLineno{align}%
\patchBothAmsMathEnvironmentsForLineno{flalign}%
\patchBothAmsMathEnvironmentsForLineno{alignat}%
\patchBothAmsMathEnvironmentsForLineno{gather}%
\patchBothAmsMathEnvironmentsForLineno{multline}%
\patchBothAmsMathEnvironmentsForLineno{eqnarray}%
}

\usepackage[pdftex,
            pdfauthor={\paperauthors},
            pdftitle={\paperasciititle},
            pdfkeywords={\paperkeywords}]{hyperref}
\usepackage{hyperxmp}
\hypersetup{
    pdfcopyright={Copyright (C) \papercopyright},
    pdflicenseurl={\paperlicenceurl}
}

\usepackage[colorinlistoftodos,textsize=scriptsize]{todonotes}

\usepackage[bottom,flushmargin,hang,multiple]{footmisc}

\usepackage[all]{hypcap} 

\usepackage{xspace} 
\usepackage{upgreek}

\def\lhcb   {\mbox{LHCb}\xspace}

\def\MagUp {\mbox{\em Mag\kern -0.05em Up}\xspace}

\ifthenelse{\boolean{uprightparticles}}%
{

 \def\Pmu         {\ensuremath{\upmu}\xspace}

 \def\Ppi         {\ensuremath{\uppi}\xspace}

 \def\Ppsi        {\ensuremath{\uppsi}\xspace}

 \def\PDelta      {\ensuremath{\Delta}\xspace}                 
 \def\PXi         {\ensuremath{\Xi}\xspace}                 
 \def\PLambda     {\ensuremath{\Lambda}\xspace}                 
 \def\PSigma      {\ensuremath{\Sigma}\xspace}                 
 \def\POmega      {\ensuremath{\Omega}\xspace}                 
 \def\PUpsilon    {\ensuremath{\Upsilon}\xspace}
 \let\oldPi\Pi
 \def\PPi         {\ensuremath{\oldPi}\xspace}

 \def\PB      {\ensuremath{\mathrm{B}}\xspace}                 
 \def\PD      {\ensuremath{\mathrm{D}}\xspace}                 
 \def\PJ      {\ensuremath{\mathrm{J}}\xspace}                 
 \def\PK      {\ensuremath{\mathrm{K}}\xspace}                 
 \def\PW      {\ensuremath{\mathrm{W}}\xspace}                 
 \def\PZ      {\ensuremath{\mathrm{Z}}\xspace}                 
 \def\Pb      {\ensuremath{\mathrm{b}}\xspace}                 
 \def\Pc      {\ensuremath{\mathrm{c}}\xspace}

 \def\Ps      {\ensuremath{\mathrm{s}}\xspace}

 \def\thebaroffset{0.0em}
}
{

 \def\Pmu         {\ensuremath{\mu}\xspace}

 \def\Ppi         {\ensuremath{\pi}\xspace}

 \def\Ppsi        {\ensuremath{\psi}\xspace}                 
                  
 \mathchardef\PDelta="7101
 \mathchardef\PXi="7104
 \mathchardef\PLambda="7103
 \mathchardef\PSigma="7106
 \mathchardef\POmega="710A
 \mathchardef\PUpsilon="7107
 \mathchardef\PPi="7105
 \def\PB      {\ensuremath{B}\xspace}                 
 \def\PD      {\ensuremath{D}\xspace}                 
 \def\PJ      {\ensuremath{J}\xspace}                 
 \def\PK      {\ensuremath{K}\xspace}                 
 \def\PW      {\ensuremath{W}\xspace}                 
 \def\PZ      {\ensuremath{Z}\xspace}                 
 \def\Pb      {\ensuremath{b}\xspace}                 
 \def\Pc      {\ensuremath{c}\xspace}

 \def\Ps      {\ensuremath{s}\xspace}

 \def\thebaroffset{0.18em}
}
\newcommand{\offsetoverline}[2][\thebaroffset]{\kern #1\overline{\kern -#1 #2}}%

\makeatletter
\ifcase \@ptsize \relax
  \newcommand{\miniscule}{\@setfontsize\miniscule{4}{5}}
\or
  \newcommand{\miniscule}{\@setfontsize\miniscule{5}{6}}
\or
  \newcommand{\miniscule}{\@setfontsize\miniscule{5}{6}}
\fi
\makeatother

\DeclareRobustCommand{\optbar}[1]{\shortstack{{\miniscule (\rule[.5ex]{1.25em}{.18mm})}
  \\ [-.7ex] $#1$}}

\def\mup        {{\ensuremath{\Pmu^+}}\xspace}
\def\mun        {{\ensuremath{\Pmu^-}}\xspace} 

\def\W      {{\ensuremath{\PW}}\xspace}

\def\Z      {{\ensuremath{\PZ}}\xspace}

\def\squark    {{\ensuremath{\Ps}}\xspace}

\def\cquark    {{\ensuremath{\Pc}}\xspace}

\def\bquark    {{\ensuremath{\Pb}}\xspace}

\def\pion   {{\ensuremath{\Ppi}}\xspace}

\def\pip    {{\ensuremath{\pion^+}}\xspace}
\def\pim    {{\ensuremath{\pion^-}}\xspace}

\def\kaon    {{\ensuremath{\PK}}\xspace}

\def\KorKbar {\kern \thebaroffset\optbar{\kern -\thebaroffset \PK}{}\xspace}

\def\Kp      {{\ensuremath{\kaon^+}}\xspace}
\def\Km      {{\ensuremath{\kaon^-}}\xspace}

\def\Kstarz  {{\ensuremath{\kaon^{*0}}}\xspace}

\def\D       {{\ensuremath{\PD}}\xspace}

\def\DorDbar {\kern \thebaroffset\optbar{\kern -\thebaroffset \PD}\xspace}
\def\Dz      {{\ensuremath{\D^0}}\xspace}

\def\Dp      {{\ensuremath{\D^+}}\xspace}
\def\Dm      {{\ensuremath{\D^-}}\xspace}

\def\DpDm    {\ensuremath{\Dp {\kern -0.16em \Dm}}\xspace}

\def\Dstarp  {{\ensuremath{\D^{*+}}}\xspace}

\def\B       {{\ensuremath{\PB}}\xspace}

\def\BorBbar {\kern \thebaroffset\optbar{\kern -\thebaroffset \PB}\xspace}
\def\Bz      {{\ensuremath{\B^0}}\xspace}

\def\Bd      {{\ensuremath{\B^0}}\xspace}

\def\BdorBdbar {\kern \thebaroffset\optbar{\kern -\thebaroffset \Bd}\xspace}
\def\Bu      {{\ensuremath{\B^+}}\xspace}

\def\Bp      {{\ensuremath{\Bu}}\xspace}

\def\Bs      {{\ensuremath{\B^0_\squark}}\xspace}

\def\BsorBsbar {\kern \thebaroffset\optbar{\kern -\thebaroffset \Bs}\xspace}

\def\jpsi     {{\ensuremath{{\PJ\mskip -3mu/\mskip -2mu\Ppsi}}}\xspace}
\def\psitwos  {{\ensuremath{\Ppsi{(2S)}}}\xspace}

\def\Y#1S{\ensuremath{\PUpsilon{(#1S)}}\xspace}

\def\LorLbar     {\kern \thebaroffset\optbar{\kern -\thebaroffset \PLambda}\xspace}

\newcommand{\decay}[2]{\ensuremath{\mathinner{#1\!\to #2}}\xspace}

\def\to                 {\ensuremath{\rightarrow}\xspace}

\def\qsq       {{\ensuremath{q^2}}\xspace}

\def\bsll     {\decay{\bquark}{\squark \ell^+ \ell^-}}

\def\AT#1     {\ensuremath{A_{\mathrm{T}}^{#1}}\xspace}           

\def\C#1      {\ensuremath{\mathcal{C}_{#1}}\xspace}                       
\def\Cp#1     {\ensuremath{\mathcal{C}_{#1}^{'}}\xspace}                    
\def\Ceff#1   {\ensuremath{\mathcal{C}_{#1}^{\mathrm{(eff)}}}\xspace}        
\def\Cpeff#1  {\ensuremath{\mathcal{C}_{#1}^{'\mathrm{(eff)}}}\xspace}       
\def\Ope#1    {\ensuremath{\mathcal{O}_{#1}}\xspace}                       
\def\Opep#1   {\ensuremath{\mathcal{O}_{#1}^{'}}\xspace}                    

\newcommand{\aunit}[1]{\ensuremath{\text{\,#1}}}       

\newcommand{\tev}{\aunit{Te\kern -0.1em V}\xspace}
\newcommand{\gev}{\aunit{Ge\kern -0.1em V}\xspace}
\newcommand{\mev}{\aunit{Me\kern -0.1em V}\xspace}
\newcommand{\kev}{\aunit{ke\kern -0.1em V}\xspace}
\newcommand{\ev}{\aunit{e\kern -0.1em V}\xspace}
 
\newcommand{\mevc}{\ensuremath{\aunit{Me\kern -0.1em V\!/}c}\xspace}
\newcommand{\gevc}{\ensuremath{\aunit{Ge\kern -0.1em V\!/}c}\xspace}
\newcommand{\mevcc}{\ensuremath{\aunit{Me\kern -0.1em V\!/}c^2}\xspace}
\newcommand{\gevcc}{\ensuremath{\aunit{Ge\kern -0.1em V\!/}c^2}\xspace}
\newcommand{\tevcc}{\ensuremath{\aunit{Te\kern -0.1em V\!/}c^2}\xspace}
\newcommand{\gevgevcccc}{\ensuremath{\gev^2\!/c^4}\xspace} 

\def\fb   {\ensuremath{\aunit{fb}}\xspace}
\def\invfb   {\ensuremath{\fb^{-1}}\xspace}

\newcommand{\stat}{\aunit{(stat)}\xspace}
\newcommand{\syst}{\aunit{(syst)}\xspace}

\def\deriv {\ensuremath{\mathrm{d}}}

\def\gsim{{~\raise.15em\hbox{$>$}\kern-.85em
          \lower.35em\hbox{$\sim$}~}\xspace}
\def\lsim{{~\raise.15em\hbox{$<$}\kern-.85em
          \lower.35em\hbox{$\sim$}~}\xspace}

\def\pt         {\ensuremath{p_{\mathrm{T}}}\xspace}

\def\evtgen     {\mbox{\textsc{EvtGen}}\xspace}

\def\geant      {\mbox{\textsc{Geant4}}\xspace}

\def\photos     {\mbox{\textsc{Photos}}\xspace}

\def\pythia     {\mbox{\textsc{Pythia}}\xspace}

\def\tell1  {TELL1\xspace}
\def\ukl1   {UKL1\xspace}

\newcommand{\lhcborcid}[1]{\href{https://orcid.org/#1}{\hspace*{0.1em}\raisebox{-0.45ex}{\includegraphics[width=1em]{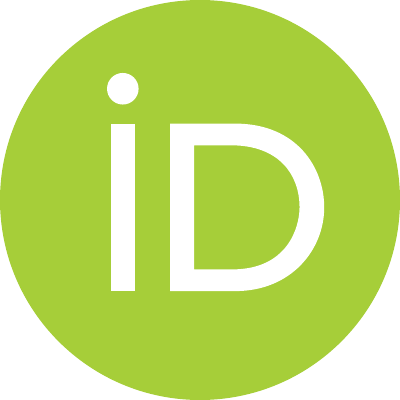}}}}

\hypersetup{
  colorlinks   = true, 
  urlcolor     = blue, 
  linkcolor    = blue, 
  citecolor    = red   
}

\ifEnableSectionTOCLinks
    \usepackage[explicit]{titlesec} 
    
    \let\oldcontentsline\contentsline
    \renewcommand

    \titleformat{\section}{\normalfont\Large\bf}{\hyperlink{tocsection.\thesection}{{\thesection} \parbox[t]{\dimexpr\textwidth-1pc}{#1}}}{1pc}{}

    \titleformat{\subsection}{\normalfont\bf}{\hyperlink{tocsubsection.\thesubsection}{{\thesubsection} \parbox[t]{\dimexpr\textwidth-1pc}{#1}}}{1pc}{}

    \titleformat{name=\section,numberless}[display]{}{}{0pt}{\normalfont\Huge\bfseries #1}
\fi

\usepackage{cite} 
\usepackage{LHCb/mciteplus}

\usepackage{longtable} 

\begin{document}

\renewcommand{\thefootnote}{\fnsymbol{footnote}}
\setcounter{footnote}{1}


\begin{titlepage}
\pagenumbering{roman}

\vspace*{-1.5cm}
\centerline{\large EUROPEAN ORGANIZATION FOR NUCLEAR RESEARCH (CERN)}
\vspace*{1.5cm}
\noindent
\begin{tabular*}{\linewidth}{lc@{\extracolsep{\fill}}r@{\extracolsep{0pt}}}
\ifthenelse{\boolean{pdflatex}}
{\vspace*{-1.5cm}\mbox{\!\!\!\includegraphics[width=.14\textwidth]{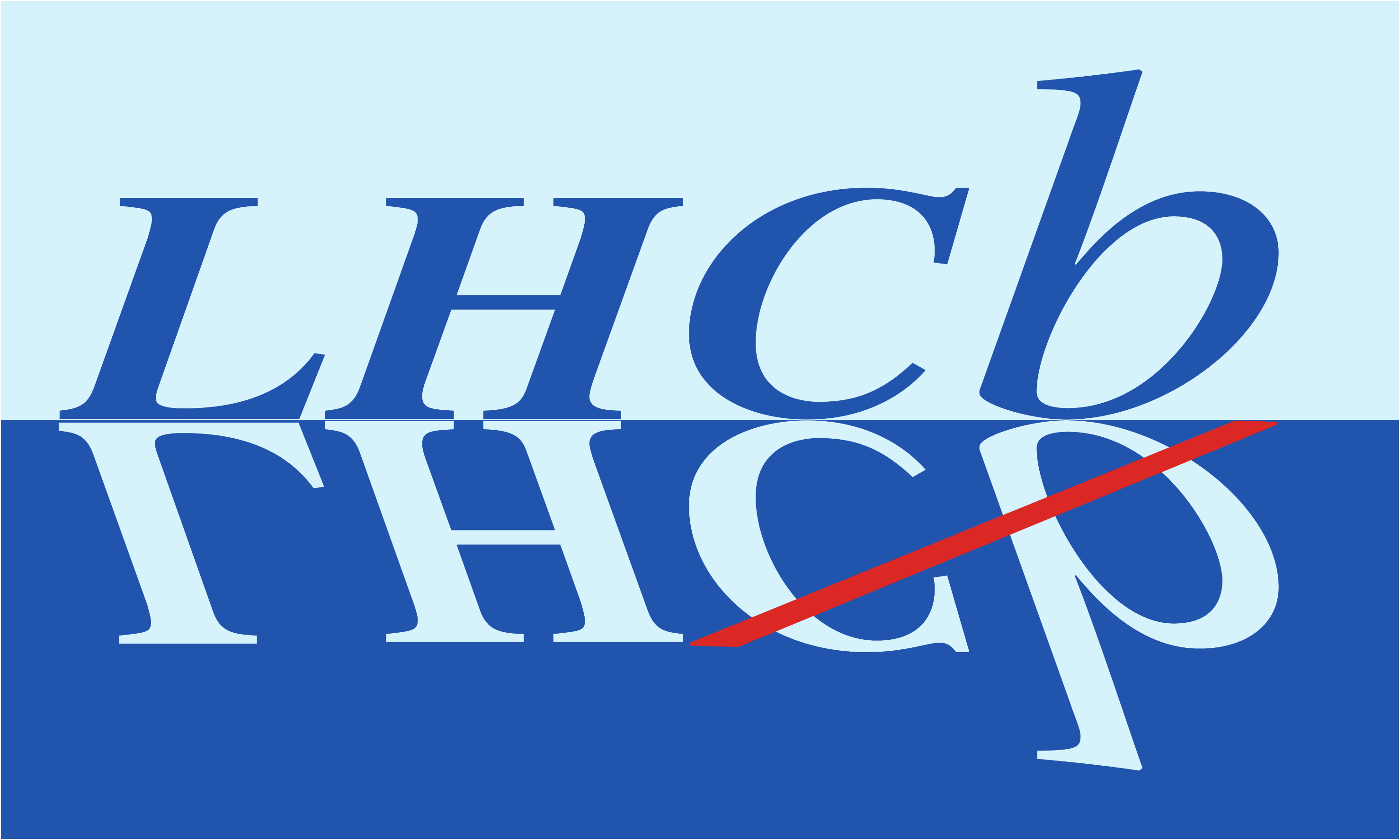}} & &}%
{\vspace*{-1.2cm}\mbox{\!\!\!\includegraphics[width=.12\textwidth]{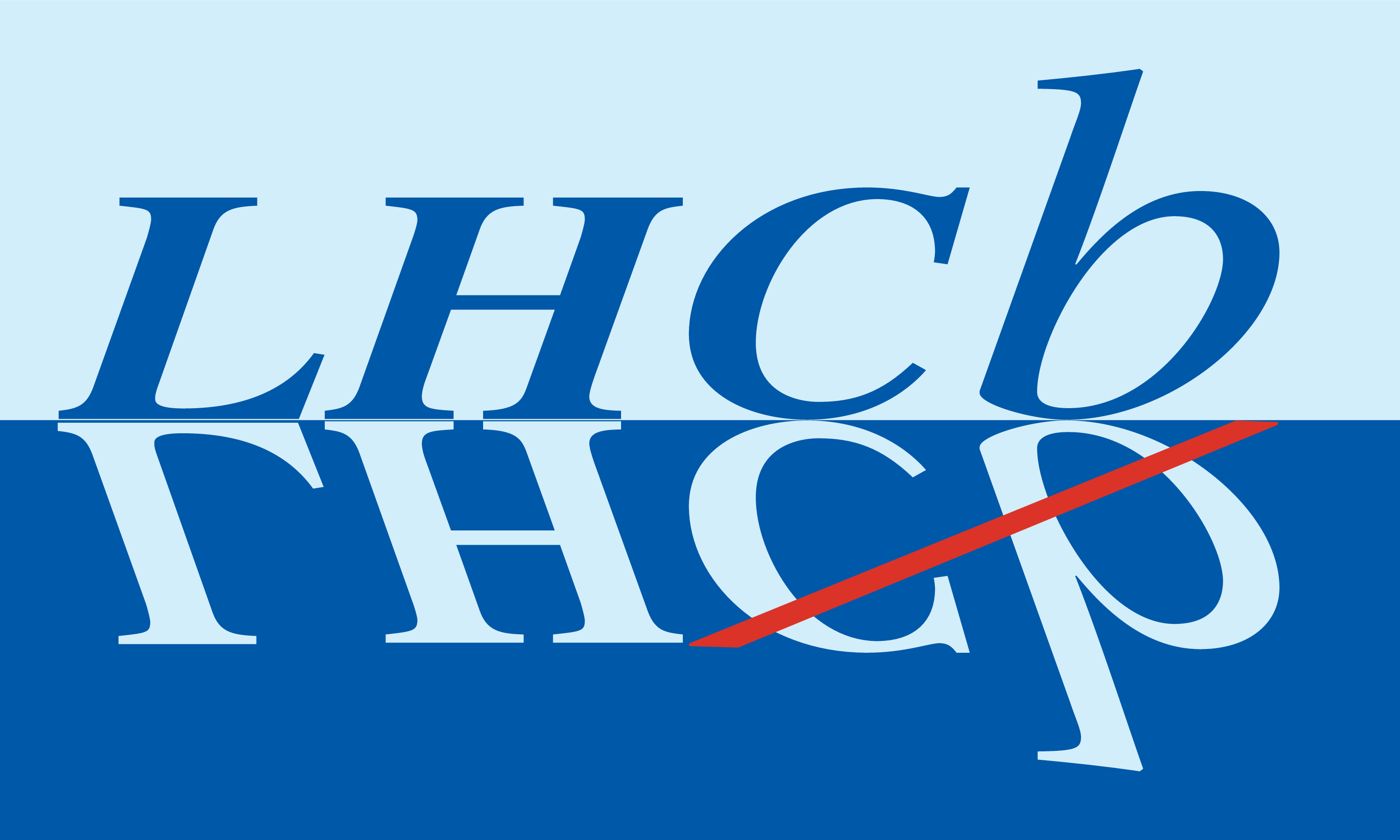}} & &}%
\\
 & & CERN-EP-2026-064 \\  
 & & LHCb-PAPER-2025-066 \\  
 & & July 31, 2026 \\ 
 & & \\
\end{tabular*}

\vspace*{4.0cm}

{\normalfont\bfseries\boldmath\huge
\begin{center}
  \papertitle 
\end{center}
}

\vspace*{2.0cm}

\begin{center}
\paperauthors\footnote{Full author list given at the end of the Letter.}
\end{center}

\vspace{\fill}

\begin{abstract}
  \noindent 
	Muon-electron universality is tested in \ensuremath{\mathinner{\Bz\!\to K^{*0}\xspace \ell^+ \ell^-}} decays, in the dilepton-invariant-mass region
	above the {{\ensuremath{\psi{(2S)}}}\xspace} resonance. The analysis uses beauty mesons produced in proton-proton collisions recorded by the LHCb detector at center-of-mass energies of
	7, 8, and 13\ensuremath{\text{\,Te\kern -0.1em V}}, corresponding to an
	integrated luminosity of 9\ensuremath{\text{\,fb}^{-1}}. The ratio of branching fractions between the muon and electron channels $R_{K^{*0}}$ is measured to be $
	1.08\,^{+0.14}_{-0.12}\text{\,(stat)}\xspace \pm 0.07\text{\,(syst)}$ for a dilepton-invariant-mass squared above 14.0\ensuremath{\text{\,Ge\kern -0.1em V}^2\!/c^4}, consistent with the standard model prediction. This result represents the most precise
	measurement of $R_{K^{*0}}$ in this region and the first such measurement performed at a hadron collider.
\end{abstract}

\vspace*{2.0cm}

\begin{center}
  Published in Phys. Rev. Lett. 137 (2026), 051803
\end{center}

\vspace{\fill}

{\footnotesize 
\centerline{\copyright~\papercopyright. \href{\paperlicenceurl}{\paperlicence}.}}
\vspace*{2mm}

\end{titlepage}


\newpage
\setcounter{page}{2}
\mbox{~}
%
%
%
%


\renewcommand{\thefootnote}{\arabic{footnote}}
\setcounter{footnote}{0}


\cleardoublepage


\pagestyle{plain} 
\setcounter{page}{1}
\pagenumbering{arabic}


Lepton flavor universality (LFU) is a key aspect of the standard model (SM), whereby the electroweak gauge interactions couple identically to the three charged lepton flavors, up to effects that depend on the lepton masses.
The flavor-changing neutral-current (FCNC) quark transition \bsll, where $\ell$ represents an electron, muon or tau lepton, could be used to precisely test LFU by comparing the rates for different lepton flavors. It occurs only through loop diagrams in the SM, leading to typical
branching fractions of $\mathcal{O}(10^{-7})$. This suppression makes these decays sensitive probes of new physics (NP), whose contributions to the decay
amplitudes can be comparable to those of the SM. Measurements of \bsll processes are sensitive to NP particles with masses up to $\mathcal{O}(50\tevcc)$, depending on the model and the nature of the couplings to SM  particles, well beyond the direct reach of current colliders~\cite{LHCb-PII-Physics}.

The universality of electroweak couplings to the three lepton families is confirmed to the level of $0.1\%$ by precise studies of SM properties, particularly
\W- and \Z-boson decays~\cite{ALEPH:2005ab,ALEPH:2013dgf}. As a consequence, SM expectations for ratios of \bsll decays involving different lepton flavors are expected to differ
from unity only through mass-suppressed effects, which are negligible for electrons and muons. As the strong interaction does not couple to leptons, hadronic uncertainties largely cancel in these ratios. This makes these quantities particularly sensitive to possible differences between transitions involving muons and electrons, providing a clean probe of NP sensitivity. 
Recent measurements of \decay{b}{s \mup \mun} processes show intriguing deviations from SM predictions in branching ratios and angular distributions, ranging
between 2 and 4 standard deviations ($\sigma$)~\cite{LHCb-PAPER-2015-051,LHCb-PAPER-2016-012,Belle:2016fev,LHCb-PAPER-2020-041,LHCb-PAPER-2020-002, LHCb-PAPER-2025-041}. However,
interpreting these measurements remains challenging due to uncertainties in hadronic form-factor
calculations~\cite{Parrott:2022zte,Gubernari:2023puw,Descotes-Genon:2023ukb} and nonperturbative effects from charm-loop
contributions~\cite{Khodjamirian:2010vf,Alguero:2023jeh,Isidori:2024lng}. Several phenomenological models offer NP interpretations to explain these
deviations~\cite{Descotes-Genon:2013wba,Altmannshofer:2014rta,London:2021lfn}. This has motivated experimentalists to investigate these anomalous
results from different perspectives, aiming to reduce theoretical uncertainties or to test similar effects in the electron decay channel through
complementary observables. In this regard, ratios between different lepton flavors provide theoretically clean tests of LFU.

Decays involving \bsll transitions are typically modeled as functions of the dilepton-invariant-mass squared \qsq. The dominant uncertainties on the ratio of muon and electron
branching fractions across the relevant \qsq range are related to the modeling of radiative effects and have been estimated to be
$\lesssim\,$1\%~\cite{Bordone:2016gaq, Isidori:2022bzw}. In contrast, NP could introduce effects to these transitions that depend on the lepton flavor, potentially at levels exceeding $20\%$ variation in the
ratios~\cite{Alguero:2022wkd,Crivellin:2022mff,Crivellin:2023saq,Buras:2024mnq,Allanach:2024jls}, while remaining consistent
with other experimental constraints. Any significant observation of LFU breaking would therefore
provide unambiguous evidence for NP. Depending on the specific nature of NP contributions, different \qsq dependence of flavor-dependent couplings could emerge, making precise tests of LFU across the \qsq phase space imperative~\cite{Hiller:2013cza,Sala:2017ihs}. 

Numerous LFU tests in \bsll transitions, focusing on electron-muon
ratios, have been performed in recent
years~\cite{LHCb-PAPER-2019-040,LHCb-PAPER-2021-004,Belle:2019oag,LHCb-PAPER-2021-038,LHCb-PAPER-2022-045,LHCb-PAPER-2022-046,CMS:2024syx,LHCb-PAPER-2024-032,LHCb-PAPER-2022-014,LHCb-PAPER-2024-056}.
Among these efforts, the most precise result to date is the combined measurement of the ratios in \decay{\Bp}{\Kp \ell^+ \ell^-} and \decay{\Bz}{\Kstarz \ell^+ \ell^-} [throughout this Letter, $\Kstarz$
indicates the $K^*(892)^0$ meson; the $\Kstarz$ is reconstructed via its dominant decay mode \decay{\Kstarz}{K^+\pi^-}] decays in a $q^2$ region  
below the \jpsi resonance ($q^2 < 6\gevgevcccc$), which validated LFU with an uncertainty of $\simeq$4\%~\cite{LHCb-PAPER-2022-046}. For more than a decade,
the \qsq region above the \psitwos resonance had been investigated only by the Belle Collaboration~\cite{Belle:2019oag}. This region has been further explored by the LHCb Collaboration with the recent measurements of the ratios in \decay{\Bs}{\phi(1020)\ell^+\ell^-}~\cite{LHCb-PAPER-2024-032} and \decay{\Bp}{\Kp \ell^+ \ell^-}~\cite{LHCb-PAPER-2024-056} decays. 

This Letter presents the first test at a $pp$ collider of muon-electron universality in \decay{\Bz}{\Kstarz \ell^+ \ell^-} decays at high dilepton invariant mass through the measurement of the branching-fraction ratio:
\begin{equation}
{\cal{R}}_{\Kstarz} \equiv \dfrac{\bigintssss_{q^2_{\mathrm{min}}}^{q^2_{\mathrm{max}}} \frac{\deriv\mathcal{B}(B^0 \to \Kstarz \mu^+ \mu^-)}{\deriv q^2} \, \deriv q^2}{\bigintssss_{q^2_{\mathrm{min}}}^{q^2_{\mathrm{max}}} \frac{\deriv \mathcal{B}(B^0 \to \Kstarz e^+ e^-)}{\deriv q^2} \, \deriv q^2},
\end{equation}
where $q^2_{\mathrm{min}}=14\gevgevcccc$ and $q^2_{\mathrm{max}}=22\gevgevcccc$ define the \qsq range of the electron and muon modes in which the analysis
is performed. The upper limit is slightly above the kinematic limit of the decay ($19.2\gevgevcccc$) as explained later. The region between $6$ and $14\gevgevcccc$ is excluded to avoid the charmonium resonances, since it is dominated by
long-distance charm-loop contributions which have shown to be consistent with the SM predictions~\cite{PDG2024}. Instead, this region is used to validate the analysis
strategy. 

The analysis uses proton-proton ($pp$) collision data recorded by the LHCb detector during the years 2011, 2012 (Run 1), and 2015--2018 (Run 2), corresponding to integrated luminosities of 1, 2, and
6\invfb at center-of-mass energies of 7, 8, and 13\tev, respectively.  Resonant \decay{\Bz}{\Kstarz \jpsi~(\to \ell^+ \ell^-)} and \decay{\Bz}{\Kstarz \psitwos~(\to \ell^+ \ell^-)} decays are also selected in \qsq ranges around the \jpsi and \psitwos mass peaks to calibrate the analysis and control systematic uncertainties related to reconstruction differences between electrons and muons. In order to cancel most detector- and trigger-related effects, ${\cal{R}}_{\Kstarz}$ is measured as a double ratio normalized by the \decay{\Bz}{\Kstarz \jpsi~(\to \ell^+ \ell^-)} control mode:
\begin{equation}
  R_{\Kstarz} = {\frac{N/\varepsilon\left(\decay{\Bz}{\Kstarz\mu^+\mu^-}\right)}{N/\varepsilon\left(\decay{\Bz}{\Kstarz\jpsi~\left(\to \mu^+\mu^-
\right)}\right)}}\bigg/{\frac{N/\varepsilon\left(\decay{\Bz}{\Kstarz e^+e^-}\right)}{N/\varepsilon\left(\decay{\Bz}{\Kstarz\jpsi~\left(\to e^+e^- \right)}\right)}},
\label{eq:RKst}
\end{equation}
where $N/\varepsilon$ is the ratio between the yield and efficiency of the corresponding signal or control mode. Residual systematic uncertainties in this double ratio may arise from distinct kinematic and radiative properties of electrons and muons.  In order to avoid experimenter’s bias, the
results of the analysis were not examined until the full procedure had been finalized.

The LHCb detector~\cite{LHCb-DP-2008-001, LHCb-DP-2014-002} is a single-arm forward spectrometer covering the pseudorapidity range $2 < \eta < 5$, designed for
the study of particles containing \bquark or \cquark quarks. 
The online event selection is performed by a trigger~\cite{LHCb-DP-2012-004}, which consists of a hardware stage based on
information from the calorimeter and muon systems, followed by a software stage which applies a full event reconstruction. At the hardware trigger stage, events
with large detector occupancy are rejected. For the muon mode, candidates are required to have a muon with high transverse momentum (\pt), and, for the electron mode, have an electron with high transverse energy in the calorimeters. The software trigger requires a two- or
three-track secondary vertex with significant displacement from any primary $pp$ interaction vertex (PV). At least one charged particle must have significant \pt and
be inconsistent with originating from a PV. A multivariate algorithm~\cite{BBDT, LHCb-PROC-2015-018} based on kinematic, geometric, and lepton
identification criteria is used for the identification of secondary vertices consistent with the decay of a \bquark hadron. 
Triggered data further undergo a centralized, offline processing step
to deliver physics-analysis-ready data across the entire \lhcb physics program~\cite{Stripping}.

Simulated samples are used to model the effects of the detector acceptance,
resolution, and the imposed selection requirements, which are needed to evaluate the reconstruction and selection efficiencies as well as to model
signal and background components.  In the simulation, $pp$ collisions are generated using \pythia~\cite{Sjostrand:2007gs,*Sjostrand:2006za} with an
\lhcb specific configuration~\cite{LHCb-PROC-2010-056}. Decays of unstable particles are described by \evtgen~\cite{Lange:2001uf}, in which final-state radiation is
generated using \photos~\cite{davidson2015photos}. The interaction of the generated particles with the detector, and its response, are implemented using the
\geant toolkit~\cite{Allison:2006ve, *Agostinelli:2002hh} as described in Ref.~\cite{LHCb-PROC-2011-006}.

The signal simulation samples are generated using differential decay rates according to a dispersion model detailed in Ref.~\cite{LHCb-PAPER-2024-011}, where local and nonlocal amplitudes are
fitted simultaneously from the decay \decay{\Bz}{\Kstarz~(\to K^+\pi^-)\mu^+\mu^-}. Separate samples are used for each data-taking period. Better consistency with data is achieved by weighting
selected signal candidates~\cite{LHCb-PAPER-2022-045} to account for known discrepancies between data and simulation in \Bz production kinematics, event occupancy, track
reconstruction, particle identification (PID), and trigger efficiencies. 

Additional corrections are applied to the simulation of the electron channel~\cite{LHCb-PAPER-2024-056}. They account for differences in \qsq resolution between data and simulation caused by bremsstrahlung emission.
The analysis is sensitive to energy losses due to the interactions of the electrons with the detector material. While the muon \qsq resolution is essentially negligible at the level of sensitivity of this analysis, the electron
resolution must be accurately modeled when computing the efficiency.

To account for residual differences between data and simulation, a smearing procedure is applied to the simulated dilepton-invariant-mass distribution
$m(e^+e^-)=\sqrt{\qsq}$ computed from the reconstructed electron momenta. The simulated distributions are determined separately for samples with $0$, $1$, or $\geq2$
 photons associated with each candidate \decay{\Bz}{\Kstarz e^+e^-} decay. For each bremsstrahlung category, probability density functions (PDFs)
extracted from simulation are  convolved with a Gaussian resolution function, and the resulting distributions are fitted to the data in the corresponding \decay{\Bz}{\Kstarz \jpsi} control mode.
The Gaussian mean and width describe the average mass shift and broadening required to align simulation with data.
The resulting corrections modify the signal efficiency by less than $4\%$.

Selection requirements are applied offline to improve signal purity. Signal candidates are reconstructed by combining two oppositely charged
electrons or muons with a $\Kstarz$ candidate, whose 
reconstructed mass is required to be within $\pm100\mev$ of the known $\Kstarz$ mass~\cite{PDG2024}. All final-state particles are required to be within
the geometric acceptance of the detectors providing particle-identification information (Cherenkov detectors, electromagnetic calorimeter, and muon detector~\cite{LHCb-DP-2014-002}) and in
the momentum range covered by the PID calibration samples. They must also form a vertex that is significantly displaced from the PV  that fits best to the flight direction of the signal \B candidate. Final-state particles are identified using a combination of multivariate classifiers that exploit information from all particle-identification subsystems, together with additional PID selection criteria that ensure tracks are consistently identified according to the relevant particle hypotheses.
To recover the \Bz mass resolution degraded by emission of bremsstrahlung photons, the electron momentum is corrected to include the energy of any reconstructed
clusters in the ECAL that match the position of the extrapolated electron track before the magnet and are not associated with any other track in the event. The
requirements apply to both signal and control modes.

Backgrounds from resonant \decay{B^0}{\Kstarz\psitwos~(\to\mu^+\mu^-)} decays contribute only marginally in the high-\qsq range.
On the contrary, the measured dielectron mass squared of the resonant decay \decay{B^0}{\Kstarz\psitwos~(\to e^+e^-)} can erroneously leak into the selected region if unrelated neutral clusters are incorrectly identified as bremsstrahlung photons
and their energy added to the energy of the electron tracks.
While such a bremsstrahlung misidentification concerns only 1\% of the candidates, the resonant decay is almost 2 orders of magnitude more abundant than the rare process of interest, and would therefore become the dominant background and significantly degrade the experimental sensitivity. For this reason the electron \qsq is always calculated with electron momenta before any
bremsstrahlung photon recovery, while the bremsstrahlung-corrected electron momenta are employed only when reconstructing the invariant mass of \Bz used for the
measurement of the ratio of interest. With these considerations, the analysis is restricted to the region $14 < \qsq < 22\gevgevcccc$ for
the measurement of $R_{\Kstarz}$, where the upper limit is extended slightly above the kinematic limit of the decay to stabilize the parametrization of the
background. Within this range, the leakage from \decay{B^0}{\Kstarz\psitwos~(\to e^+e^-)} decays is negligible.

To further reduce background contamination from combinations of unrelated tracks (combinatorial background) and misreconstructed decays due to missed or
misidentified particles, two sets of optimizations are performed. Combinatorial background is suppressed using two distinct multivariate classifiers for the
muon and electron modes, which share kinematic, vertex, and geometric features of the signal candidate as input. The classifiers are adapted from those in
Ref.~\cite{LHCb-PAPER-2022-045}, with the working points of both modes optimized separately. They are trained using the simulated signal samples as signal proxy, and candidates with both leptons of the same charge sign in data (the same-sign sample) as background proxy. Backgrounds from specific misidentified decays are suppressed with dedicated mass vetoes and tight PID requirements~\cite{LHCb-PAPER-2022-045}. Residual misidentified backgrounds are modeled in the fit.

The lepton-universality ratios are extracted from simultaneous unbinned maximum-likelihood fits to the \Bz invariant-mass distributions of \decay{\Bz}{\Kstarz\ell^+\ell^-} candidates. The
fits incorporate the measured efficiency values according to Eq.(~\ref{eq:RKst}) and are performed jointly for the muon and electron modes, as well as across the high-\qsq
(signal) and $\jpsi$ (normalization) regions. The analysis strategy is first validated by measuring $r_{J/\psi}$, the ratio of the branching fractions $\decay{B^0}{\Kstarz \jpsi~(\to \mu^+\mu^-)}$ to $\decay{B^0}{\Kstarz \jpsi~(\to e^+e^-)}$. For this measurement, the selected range is $9.0<\qsq< 10.2\gevgevcccc$ for the muon mode and $7.6<\qsq< 9.8\gevgevcccc$ for the electron mode. In addition, $R_{\psitwos}$, defined analogously to $R_{\Kstarz}$ using the $\decay{B^0}{\Kstarz \psitwos~(\to \ell^+\ell^-)}$ decay, is validated in the range $12.9< \qsq< 14.0\gevgevcccc$ for the muon mode and $11.4<\qsq< 14.0\gevgevcccc$ for the electron mode.
The \qsq region definitions here guarantee a consistent selection efficiency between the
control and signal modes~\cite{LHCb-PAPER-2024-056}, rather than consistency between the muon and electron modes.
In these validation studies, the \Bz mass resolution of the control modes is improved by constraining the dilepton invariant mass to that of the \jpsi or \psitwos
meson~\cite{PDG2024}. The fit configuration matches that used in Ref.~\cite{LHCb-PAPER-2022-045}, yielding measured values of $r_{J/\psi} = 1.035 \pm
0.013\pm0.035$ and $R_{\psitwos} = 1.034 \pm 0.033 \pm 0.012$, where the first uncertainty is statistical and the second is from efficiency-related systematic. 

In the fit in the high-\qsq region,  weights are applied to both the data and simulated samples of the muon channel, 
to reduce differences in the selection efficiency of the electron and muon modes which are \qsq dependent.
This reduces bias in the efficiency ratio caused by an imperfect modeling of
the form factors used to generate the simulated samples. The weight is defined as $w_{\mathrm{eff}}(q^2) \equiv \varepsilon^{ee}(\qsq) / \varepsilon^{\mu\mu}(\qsq)$, where $\varepsilon^{ee}(\qsq)$ and
$\varepsilon^{\mu\mu}(\qsq)$ are the efficiencies of the electron and muon modes, respectively, obtained from efficiency-corrected signal simulation
samples~\cite{LHCb-PAPER-2024-056}. The muon channel is chosen for reweighting owing to its higher statistical precision and better \qsq resolution. The weights
are evaluated in \qsq intervals of $0.2\gevgevcccc$ separately for each data-taking period, with statistical uncertainties evaluated with a bootstrapping
method.

The \Bz-candidate mass distribution is modeled separately for the electron and muon channels. For the muon channel, the signal shape is described by the sum
of a modified Crystal Ball function~\cite{Skwarnicki:1986xj} with tails on both sides of the peak (double-sided Crystal Ball) and a Gaussian function sharing the same peak position, with all shape parameters fixed from simulation. For the electron channel, the
model parameters depend on the number of bremsstrahlung photons associated with the electron tracks. Each category is described by a double-sided Crystal Ball function, with shape parameters and relative fractions fixed from simulation. To account for minor data-simulation discrepancies, the peak position is shifted and its width scaled by parameters allowed to vary in the fit and shared between the control and signal modes, in order to maximize statistical power and reduce systematic uncertainties.

Backgrounds are modeled with dedicated PDFs for each significant component. The combinatorial background is parametrized with the empirical function $f(m) = \left[ 1+(2^\beta-1)e^{- \frac{m-\alpha}{\sigma} }\right]^{-1/\beta}$, where $m$ is the $K\pi\ell\ell$ invariant mass and $\alpha$, $\beta$, and $\sigma$ are free parameters. The initial parameter values are determined from the same-sign sample and
are subsequently allowed to vary within their uncertainties in the fit. Another source of background is due to partially reconstructed \decay{\Bp}{\Kp \pip \pim
\ell^+ \ell^-} decays where the $\pip$ is missing. It is modeled using weighted simulation samples to account for the contributions of the observed intermediate resonances~\cite{LHCb-PAPER-2024-046}, ensuring an accurate description of
its line shape. 
A distinctive background in the electron channel originates from hadrons misidentified as electrons, for which the PID performance differs between data and
simulation.  Its shape is obtained using a data-driven procedure, owing to the complex composition of the background (including contributions from the processes \decay{\Bz}{\Kstarz\pip\pim} and $b$-meson semileptonic decays). Calibration samples of $\pi^{\pm}$ and $K^{\pm}$ mesons are selected by reconstructing
\decay{\Dstarp}{\Dz~(\to \Km\pip)\pip} decays in data~\cite{LHCb-PUB-2016-021, LHCb-DP-2018-001}, from which transfer functions are derived as the ratios of the probabilities for these hadrons to
pass or fail the electron-identification criteria. Candidate \decay{\Bz}{\Kstarz h^{+}h^{-}} decays, with the final-state hadrons $h$ required to fail
the baseline electron-identification requirements, are then used to model the misidentified background. The expected yields and distributions of candidates passing the electron-identification criteria are
obtained by convolving the transfer functions with this sample. These data-driven yield estimates are included in the fit as external constraints~\cite{LHCb-PAPER-2022-045}.

\begin{figure}[tpb]
	\centering
	\includegraphics[width=\linewidth]{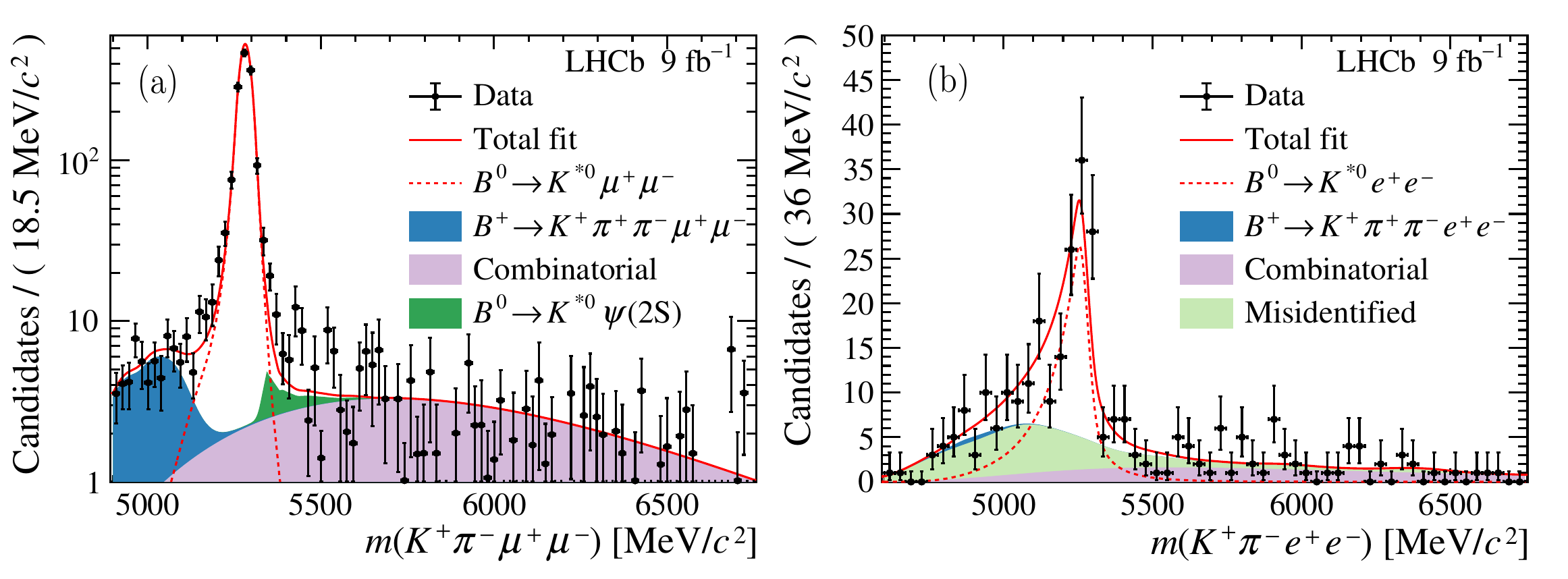}
	\caption{Invariant-mass distributions of the \decay{\Bz}{\Kstarz \ell^+ \ell^-} candidates in the high-\qsq region for the (a) muon and
		(b) electron channel with the fit results also shown. The left-side cutoff of the combinatorial background distribution arises from the kinematic boundaries of the accessible phase space. 
    }%
	\label{fig:./FINALF}
\end{figure}

Figure~\ref{fig:./FINALF} shows the fit to the \Bz invariant-mass distributions of the \decay{\Bz}{\Kstarz\ell^+\ell^-} candidates in the high-\qsq region. Good agreement between the fit model and data is observed for both lepton flavors. The fitted signal yields are $125_{-16}^{+14}$ and $1408 \pm 38$ for the electron and muon channels, respectively; the uncertainties are statistical only.

Systematic uncertainties are evaluated and categorized into two primary sources: those related to efficiency corrections and those affecting the yield determination. Efficiency-related uncertainties are assessed by testing alternative correction methods or varying the weighting and smearing parameters within their
uncertainties~\cite{LHCb-PAPER-2022-045}. Key terms include particle identification, tracking efficiency, event multiplicity, and trigger efficiencies, each
contributing 0.1--0.8\% to the uncertainty on the measured ratios. The overall uncertainty related to efficiencies sums each term in quadrature, and is 1.2\%.
Since $R_{\Kstarz}$ is expected to be independent of the kinematic and topological variables after efficiency correction, its residual dependence on these
variables  is evaluated using the control mode~\cite{LHCb-PAPER-2022-045}. The model dependence of the signal efficiency is assessed using alternative form-factor models in simulation~\cite{Ali:1999mm},
where the cancellation achieved via the implementation of $w_{\mathrm{eff}}(q^2)$ reduces the systematic uncertainty to $\sim2.0$\%. The \qsq smearing uncertainty accounting for the resolution on the dilepton invariant mass is evaluated by varying the smearing parameters and employing alternative fit models. The uncertainty on the integrated luminosity of each data-taking period is propagated to the final efficiency calculation through a bootstrapping procedure.

Fit-related uncertainties are evaluated using an ensemble of pseudoexperiments. Samples are generated from the baseline fit model, with
signal and background yields fluctuated according to Poisson distributions centered on their fitted values. For each source of systematic uncertainty, an alternative fit model is tested. Each generated sample is fitted twice: once with the baseline model and once with the alternative model. The corresponding systematic uncertainty is determined by summing in quadrature the standard deviation of the resulting pull distribution. This method directly translates model variations into an uncertainty on $R_{K^{*0}}$.
The dominant uncertainty arises from the electron channel, specifically from the data-driven estimation of the misidentified background. The systematic uncertainty
associated with this method is assessed by altering the particle-identification criteria in the pion and kaon calibration samples~\cite{LHCb-PUB-2016-021, LHCb-DP-2018-001}.  
Variations in the signal and combinatorial background shapes, along with the inclusion of the partially reconstructed background, are also propagated through the pseudoexperiment-based procedure. A summary of systematic uncertainties is provided in Table~\ref{tab:effsystematic}.

\begin{table}[tpb]
	\centering
	\caption{Relative systematic uncertainties on $R_{\Kstarz}$. For reference, the statistical uncertainty is $12.0\%$.}
	\label{tab:effsystematic}
	\begin{tabular}{lc}
\hline
	 Source & Uncertainty (\%) \\
	\hline
	 Efficiency correction &  1.2\\
	 Kinematic and topological dependence & 0.7\\
	 Model dependence & 2.0\\
	 \qsq smearing & 0.9\\
	 Integrated luminosity & 0.9\\
	 \hline
Misidentified background & 5.2 \\
Signal line shape & 1.7 \\
Combinatorial background & 1.3 \\
Partially reconstructed background & 2.0 \\
\hline
	\end{tabular}
\end{table}

 The log-likelihood curve from the baseline fit is shown in Fig.~\ref{fig:./LL_scan_comparison} in the End Matter, along with the log-likelihood curve convolved with a Gaussian function to account for systematic uncertainties, which are considered to be uncorrelated and combined in quadrature.
The measured value of $R_{\Kstarz}$ in the high-\qsq region is 
\begin{equation*}
R_{\Kstarz} = 1.08^{\ +0.14}_{\ -0.12}\stat \pm 0.07\syst,
\end{equation*}
where the statistical uncertainty is derived from the profile likelihood scan without systematic uncertainties, and the systematic uncertainty is obtained from
the difference in quadrature between the total and statistical uncertainties. The result is consistent with the SM prediction within approximately 0.5$\sigma$, but it can exclude certain new physics models (for example, those involving $Z^{\prime}$-boson exchange~\cite{MunirBhutta:2020ber}) at the 2$\sigma$ level.

In summary,  the first hadron-collider measurement of the ratio of branching ratios of \decay{\Bz}{\Kstarz e^+e^- (\mu^+\mu^-)}
decays is performed in the high-\qsq region using the LHCb Run~1 and Run~2 dataset, corresponding to an integrated luminosity of 9\invfb. The result is $R_{\Kstarz} = 1.08\,^{+0.14}_{-0.12}\stat \pm 0.07\syst$, which is consistent with the previous result reported by the Belle Collaboration~\cite{Belle:2019oag} and achieves a threefold improvement in
precision. The present Letter therefore provides the most precise test to date of the LFU ratio $R_{\Kstarz}$ in the high-\qsq region.  
A summary of the results of $R_{\Kstarz}$ from this work, the LHCb measurement in the lower-\qsq region~\cite{LHCb-PAPER-2022-046}, and the SM prediction is presented in Fig.~\ref{fig:comparison} in the End Matter. All existing measurements of $R_{\Kstarz}$ are in agreement with
the SM. 
These findings provide a stringent test of LFU in \bsll transitions at high-\qsq, complementing early studies at lower dilepton invariant masses.
The full exploitation of larger future datasets, for which the precision is expected to improve by around a factor of 2~\cite{LHCb-DP-2022-002}, will be essential for clarifying the observations and probing potential new physics effects in FCNC processes.

\section*{Acknowledgements}
%
%
\noindent We express our gratitude to our colleagues in the CERN
accelerator departments for the excellent performance of the LHC. We
thank the technical and administrative staff at the LHCb
institutes.
We acknowledge support from CERN and from the national agencies:
ARC (Australia);
CAPES, CNPq, FAPERJ and FINEP (Brazil); 
MOST and NSFC (China); 
CNRS/IN2P3 and CEA (France);  
BMFTR, DFG and MPG (Germany);
INFN (Italy); 
NWO (Netherlands); 
MNiSW and NCN (Poland); 
MEC/IFA (Romania); 
MICIU and AEI (Spain);
SNSF and SER (Switzerland); 
NASU (Ukraine); 
STFC (United Kingdom); 
DOE NP and NSF (USA).
We acknowledge the computing resources that are provided by ARDC (Australia), 
CBPF (Brazil),
CERN, 
IHEP and LZU (China),
IN2P3 (France), 
KIT and DESY (Germany), 
INFN (Italy), 
SURF (Netherlands),
Polish WLCG (Poland),
IFIN-HH (Romania), 
PIC (Spain), CSCS (Switzerland), 
GridPP (United Kingdom),
and NSF (USA).  
We are indebted to the communities behind the multiple open-source
software packages on which we depend.
Individual groups or members have received support from
RTP (Australia), 
Key Research Program of Frontier Sciences of CAS, CAS PIFI, CAS CCEPP (China); 
Minciencias (Colombia);
EPLANET, Marie Sk\l{}odowska-Curie Actions, ERC and NextGenerationEU (European Union);
A*MIDEX, ANR, IPhU and Labex P2IO, and R\'{e}gion Auvergne-Rh\^{o}ne-Alpes (France);
Alexander-von-Humboldt Foundation (Germany);
ICSC (Italy); 
Severo Ochoa and Mar\'ia de Maeztu Units of Excellence, GVA, XuntaGal, GENCAT, InTalent-Inditex and Prog.~Atracci\'on Talento CM (Spain);
the Leverhulme Trust, the Royal Society and UKRI (United Kingdom).




\section*{End Matter}
\label{sec:end_matter}
\begin{figure}[hpb]
	\centering
	\includegraphics[width=0.7\linewidth]{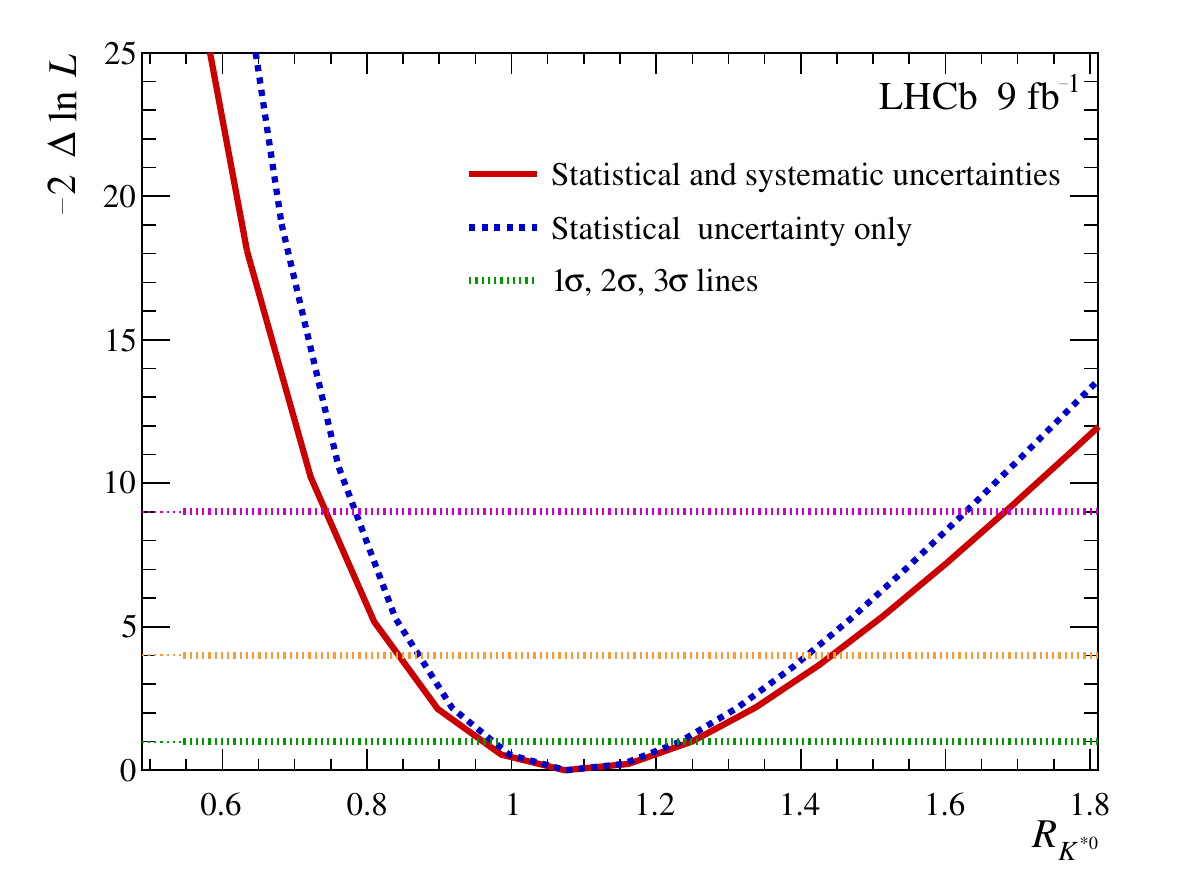}
	\caption{Log-likelihood profile of $R_{\Kstarz}$ with (red solid) and without (blue dashed) systematic uncertainties. 
    }
	\label{fig:./LL_scan_comparison}
\end{figure}

The likelihood profile of $R_{\Kstarz}$ is evaluated both with and without the inclusion of systematic uncertainties in order to assess their impact.
To construct the profile, a series of unbinned maximum-likelihood fits is performed, each with $R_{\Kstarz}$ fixed to a given value,
while all other nuisance parameters are profiled. The statistical-only likelihood profile is first obtained from these fits.
Systematic uncertainties are then incorporated by convolving the statistical-only likelihood with Gaussian functions,
where the width of each Gaussian corresponds to the size of the associated systematic uncertainty.  The resulting likelihood profiles showing
the impact of systematic uncertainties on the measurement are presented in Fig.~\ref{fig:./LL_scan_comparison}.

\begin{figure}[hpb]
	\centering
	\includegraphics[width=0.7\linewidth]{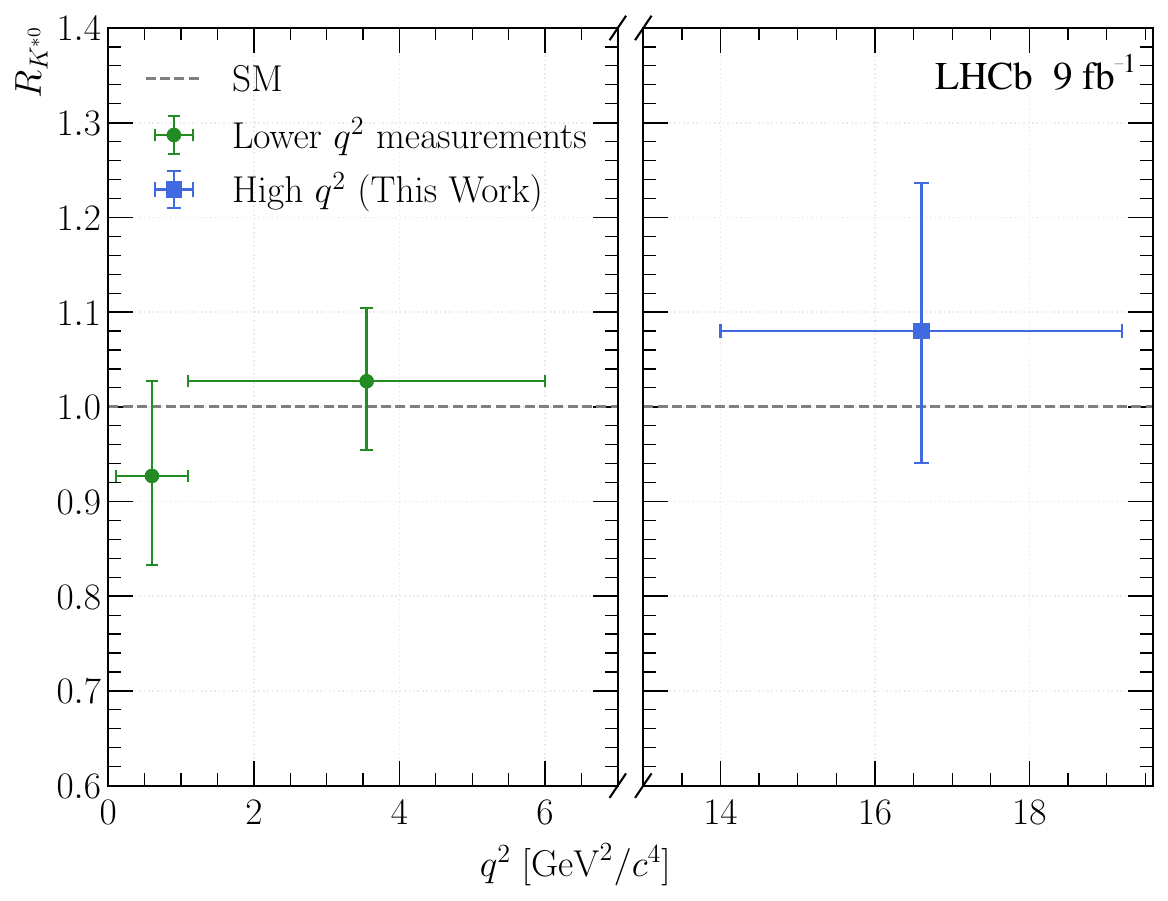}
	\caption{ Summary of the measurement of $R_{\Kstarz}$ from this work (blue square point), the LHCb measurements in the lower-\qsq region~\cite{LHCb-PAPER-2022-046} (green
		dot points) and the SM prediction (dashed line). The vertical error bars
	represent the total uncertainties.}
	\label{fig:comparison}
\end{figure}

Figure~\ref{fig:comparison} shows a summary of the results of $R_{\Kstarz}$ from this work, the LHCb measurement in the lower \qsq region~\cite{LHCb-PAPER-2022-046} and the SM prediction.

\clearpage

\addcontentsline{toc}{section}{References}
\bibliographystyle{LHCb/LHCb}
\bibliography{LHCb/main,LHCb/standard,LHCb/LHCb-PAPER,LHCb/LHCb-CONF,LHCb/LHCb-DP,LHCb/LHCb-TDR}

\newpage
\centerline
{\large\bf LHCb collaboration}
\begin
{flushleft}
\small
R.~Aaij$^{38}$\lhcborcid{0000-0003-0533-1952},
M. ~Abdelfatah$^{69}$,
A.S.W.~Abdelmotteleb$^{57}$\lhcborcid{0000-0001-7905-0542},
C.~Abellan~Beteta$^{51}$\lhcborcid{0009-0009-0869-6798},
F.~Abudin\'en$^{59}$\lhcborcid{0000-0002-6737-3528},
T.~Ackernley$^{61}$\lhcborcid{0000-0002-5951-3498},
A. A. ~Adefisoye$^{69}$\lhcborcid{0000-0003-2448-1550},
B.~Adeva$^{47}$\lhcborcid{0000-0001-9756-3712},
M.~Adinolfi$^{55}$\lhcborcid{0000-0002-1326-1264},
P.~Adlarson$^{87}$\lhcborcid{0000-0001-6280-3851}
% ...

{\footnotesize \it

% ...
$^{38}$Nikhef National Institute for Subatomic Physics, Amsterdam, Netherlands\\
% ...
$^{47}$Instituto Galego de F{\'\i}sica de Altas Enerx{\'\i}as (IGFAE), Universidade de Santiago de Compostela, Santiago de Compostela, Spain\\
% ...
$^{51}$Physik-Institut, Universit{\"a}t Z{\"u}rich, Z{\"u}rich, Switzerland\\
% ...
$^{55}$H.H. Wills Physics Laboratory, University of Bristol, Bristol, United Kingdom\\
% ...
$^{57}$Department of Physics, University of Warwick, Coventry, United Kingdom\\
% ...
$^{59}$School of Physics and Astronomy, University of Edinburgh, Edinburgh, United Kingdom\\
$^{60}$School of Physics and Astronomy, University of Glasgow, Glasgow, United Kingdom\\
$^{61}$Oliver Lodge Laboratory, University of Liverpool, Liverpool, United Kingdom\\
% ...
$^{69}$Syracuse University, Syracuse, NY, United States\\
% ...
$^{87}$Department of Physics and Astronomy, Uppsala University, Uppsala, Sweden, associated to $^{60}$\\
% ...
}
\end{flushleft}

\end{document}